\documentclass{emulateapj}

\slugcomment{To appear in the {\em Astrophysical Journal}}
\shorttitle{Emission from Quiescent Massive Black Holes}
\shortauthors{Wrobel, Terashima \& Ho}
\begin{document}
\title{Outflow-Dominated Emission from the Quiescent Massive Black Holes
       in NGC\,4621 and NGC\,4697}
\author{J. M. Wrobel,\altaffilmark{1}
        Y. Terashima,\altaffilmark{2}
        and L. C. Ho\altaffilmark{3}}
\altaffiltext{1}{National Radio Astronomy Observatory, P.O. Box O,
Socorro, NM 87801; jwrobel@nrao.edu}
\altaffiltext{2}{Department of Physics, Faculty of Science, 
Ehime University, Matsuyama 790-8577, Japan;
terasima@astro.phys.sci.ehime-u.ac.jp}
\altaffiltext{3}{The Observatories of the Carnegie Institution of
Washington, 813 Santa Barbara Street, Pasadena, CA 91101;
lho@ociw.edu}

\begin{abstract}
The nearby elliptical galaxies NGC\,4621 and NGC\,4697 each host a
supermassive black hole with $M_\bullet > 10^8 M_\odot$.  Analysis of
archival {\em Chandra\/} data and new NRAO Very Large Array data shows
that each galaxy contains a low-luminosity active galactic nucleus
(LLAGN), identified as a faint, hard X-ray source that is
astrometrically coincident with a faint 8.5-GHz source.  The latter
has a diameter less that 0.3\arcsec\, (26~pc for NGC\,4621, 17~pc for
NGC\,4697).  The black holes energizing these LLAGNs have Eddington
ratios $L(2-10~keV) / L(Edd) \sim 10^{-9}$, placing them in the
so-called quiescent regime.  The emission from these quiescent black
holes is radio-loud, with $log~R_X = log~\nu L_{\nu}(8.5~GHz) /
L(2-10~keV) \sim -2$, suggesting the presence of a radio outflow.
Also, application of the radio--X-ray--mass relation from Yuan \& Cui
for quiescent black holes predicts the observed radio luminosities
$\nu L_{\nu}(8.5~GHz)$ to within a factor of a few.  Significantly,
that relation invokes X-ray emission from the outflow rather than from
an accretion flow.  The faint, but detectable, emission from these two
massive black holes is therefore consistent with being
outflow-dominated.  Observational tests of this finding are suggested.
\end{abstract}

\keywords{galaxies: active ---
          galaxies: individual (NGC\,4621, NGC\,4697) ---
          galaxies: nuclei ---
          radio continuum: galaxies --- X-rays: galaxies}

\section{Motivation}

Dynamical studies have established that supermassive black holes, with
masses $M_\bullet \sim 10^6 - 10^9 M_\odot$, occur in the nuclei of
most nearby galaxies with stellar bulges \citep[e.g.,][]{kor04}.  Yet
few of these massive black holes are observed as luminous active
galactic nuclei (AGNs).  Rather, the Palomar spectroscopic survey by
\citet{ho97} showed that the majority have either no AGN signatures or
only the weak AGN signatures that define them as low-luminosity AGNs
(LLAGNs; $L(H\alpha) \le 10^{40}$~ergs~s$^{-1}$).  Also, Chandra
surveys of selected Palomar LLAGNs \citep{ho01,ter03} commonly find
X-ray nuclei with $L(2-10~keV) = 10^{38} - 10^{42}$~ergs~s$^{-1}$,
leading to $L(2-10~keV) < 10^{42}$~ergs~s$^{-1}$ as an X-ray
definition of a LLAGN.  For the black hole masses involved, such
H$\alpha$ and X-ray luminosities are highly sub-Eddington, that is,
much less than the Eddington luminosity $L(Edd) = 1.3 \times 10^{38}
(M_\bullet / M_\odot )$~ergs~s$^{-1}$.  Understanding the radiative
quiescence of these massive black holes has important implications for
accretion physics, fuelling and feedback mechanisms, and black-hole
growth over cosmic time \citep{ho04,pel05}.

Early theoretical models for weakly-radiating massive black holes
invoked radiatively-inefficient inflows \citep[e.g.,][]{nar95},
jets/outflows \citep[e.g.,][]{fal00}, and combinations of the two
\citep[e.g.,][]{yua02a}.  The models predicted that continuum emission
could emerge in the radio and hard X-ray regions.  For LLAGNs, these
frequency regions offer strong tests of the models because they
benefit from high contrast against the stellar emission, little or no
obscuration, and subarcsecond angular resolution and astrometry.
Significantly, a study of Palomar LLAGNs using {\em Chandra\/}, the
Very Large Array (VLA), and the Very Long Baseline Array (VLBA), found
that the majority are radio loud, defined as $log~R_X = log~\nu
L_{\nu}(5~GHz) / L(2-10~keV) = -4.5$ or higher \citep{ter03}.  Since
LLAGNs may possess radiatively-inefficient accretion flows
\citep[e.g.,][]{yua02b} and radio continuum emission is
outflow-dominated \citep[e.g.,][]{fal99,nag05}, this suggests that
such accretion flows can produce outflows more efficiently than
standard geometrically thin accretion disks \citep{liv99,mei01}.
Moreover, such radio continuum outflows might have sufficiently stable
``astrometric footprints'' to constrain the proper motions of their
galaxy hosts in the extragalactic frame \citep[e.g.,][]{bie00},
especially for galaxy hosts which lack interstellar water masers.

Building upon the work by \citet{ter03}, \citet{ho03b} and
\citet{ter05} searched in the radio and X-ray regions for LLAGN
signatures from the weakest accretors, namely nearby Palomar nuclei
showing no or very weak $H\alpha$ \citep{ho03a}.  Those targets were
mainly disk galaxies and the new {\em Chandra\/} detections of them
correspond to X-ray nuclei with $L(2-10~keV) < 10^{38}$~ergs~s$^{-1}$
\citep{ho03b,ter05}.  With their typical $M_\bullet < 10^8 M_\odot$,
these accretors are feeble emitters with Eddington ratios $L(2-10~keV)
/ L(Edd) < 10^{-7}$.  Still, \citet{ter05} obtained VLA detections of
radio nuclei at the levels predicted from the radio-loudness relation
\citep{ter03}, proving that the trait of radio loudness holds below
$L(2-10~keV) = 10^{38}$~ergs~s$^{-1}$ and continuing to emphasize the
potential importance of radio outflows from these modest-mass black
holes.

Encouraged by these X-ray/radio trends, a next step is to search for
continuum signatures of LLAGNs in nearby elliptical galaxies, with
their typically higher $M_\bullet$.  Archival and published data from
{\em Chandra\/} and the VLA are used in \S~\ref{data-old} to identify
two candidate LLAGNs in nearby elliptical galaxies.  New
high-resolution observations of these candidates with the
VLA\footnote{Operated by the National Radio Astronomy Observatory,
which is a facility of the National Science Foundation, operated under
cooperative agreement by Associated Universities, Inc.}  \citep{tho80}
are reported in \S~\ref{data-new}.  The implications of the new VLA
imaging are explored in \S~\ref{astrometry} regarding the astrometry,
and in \S~\ref{photometry} regarding the photometry.  \S~\ref{summary}
closes with a summary of this work and suggestions for future
directions.

\section{Prior Data}\label{data-old}

A Palomar elliptical galaxy \object{NGC\,4621} and a southern
elliptical galaxy \object{NGC\,4697} were selected.  NGC\,4621 is an
absorption-line nucleus with an H$\alpha$ upper limit from
\citet{ho03a}.  NGC\,4697 has modest LINER characteristics (Pinkney et
al. 2005; J. Pinkney, priv.\ comm.).  For each galaxy, Table~1 lists
its surface-brightness-fluctuation distance and scale, as well the
mass and Eddington luminosity of its black hole.  Table~2 gives the
centroid positions of the galaxies from NED/2MASS.

\placetable{tab1}

\placetable{tab2}

\subsection{VLA}

At the NED/2MASS centroid positions, the galaxies are neither confused
nor detected at 1.4~GHz, and thus are less than 2.5~mJy at 45\arcsec\,
resolution \citep{con98} and less than 1~mJy at 5\arcsec\, resolution
\citep{whi97}.  From VLA data at 5~GHz and 5\arcsec\, resolution,
NGC\,4621 is less than 0.5~mJy \citep{wro91} and NGC\,4697 is less
than 0.6~mJy \citep{bir85}.  From VLA data at 8.5~GHz and 10\arcsec\,
resolution, NGC\,4621 has a candidate LLAGN that is unresolved and has
a low-resolution flux density of 0.153$\pm$0.014~mJy \citep{wro00b}.
From VLA data at 8.5~GHz and about 3\arcsec\, resolution,
\citet{kra02} reported an upper limit for NGC\,4697; but the
sensitivity values claimed were not plausable given the exposure time
so the VLA archival data were analyzed following the strategies
outlined in \S~\ref{data-new} and leading to a 4 $\sigma$ upper limit
of 0.164~mJy for these low-resolution data.

\subsection{Chandra}

X-ray data for the galaxies were retrieved from the {\em Chandra\/}
archives.  These galaxies were observed with the ACIS-S3
back-illuminated CCD chip.  The data were reprocessed and then
analyzed with the CIAO version 3.2 software package.  Background
levels were low and stable.  The UT observation date and net exposure
times appear in Table~2.  Corrections were made for known aspect
offsets\footnote{
http://cxc.harvard.edu/cal/ASPECT/fix\_offset/fix\_offset.cgi}.

Figures~1 and 2 show the resulting images in the full (0.5-8~keV),
soft (0.5-2~keV) and hard (2-8~keV) energy ranges.  Several X-ray
sources are seen in the central regions of each galaxy.  This work
focuses on the nearest and brightest X-ray sources to the NED/2MASS
centroid positions marked in Figures~1 and 2.  For those X-ray
sources, the WAVDETECT tool was used to measure their positions in the
full energy range; the parameters used in the detection procedure are
given in \citet{ter03}.  The X-ray positions of these candidate LLAGNs
are given in Table~2 along with the diameter of the position error
circle, 1.2\arcsec\, at the 90\% confidence limit\footnote{
http://cxc.harvard.edu/cal/ASPECT/celmon}.  These X-ray positions are
also marked in Figures~1 and 2.

\begin{figure}
\plotone{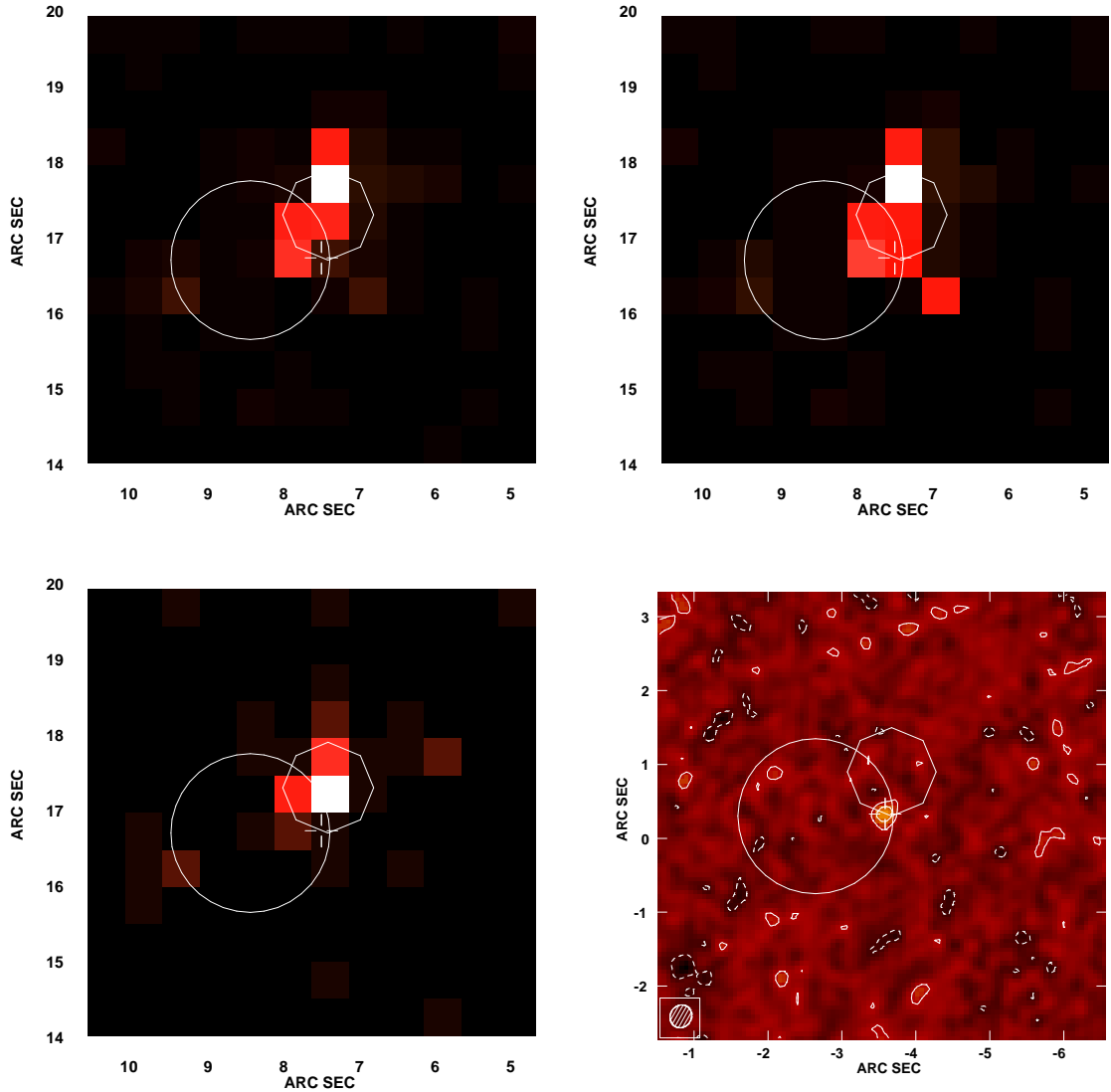}
\caption{Images of the central 6\arcsec\, (530 pc) of NGC\,4621.  The
symbols mark positions and their errors at the 90\% confidence level.
The NED/2MASS position of the galaxy centroid is marked with a circle,
while the octagon marks the {\em Chandra\/} position of the nearest
and brightest source to the NED/2MASS position.  The high-resolution
VLA position from this work is marked with a cross with a gap.  {\em
Top left:\/} {\em Chandra\/} image at 0.5-8 keV with a peak of 16
counts.  {\em Top right:\/} {\em Chandra\/} image at 0.5-2 keV with a
peak of 12 counts.  {\em Bottom left:\/} {\em Chandra\/} image at 2-8
keV with a peak of 5 counts.  In the {\em Chandra\/} images, the
coordinate labels are relative to the aim point (to convey that
astrometric distortions should be negligible) and the logarithmic
color scale spans zero to the peak.  {\em Bottom right:\/} VLA image
of Stokes $I\/$ emission at a frequency of 8.460~GHz.  Coordinate
labels are relative to the pointing position.  Natural weighting was
used, giving an rms noise of 0.018~mJy~beam$^{-1}$ (1 $\sigma$) and
beam dimensions at FWHM of 0.32\arcsec\, $\times$ 0.29\arcsec\, with
elongation PA = -30\arcdeg\, (hatched ellipse).  Contours are at -6,
-4, -2, 2, 4, and 6 times 1 $\sigma$.  Negative contours are dashed
and positive ones are solid.  Linear color scale spans
-0.06~mJy~beam$^{-1}$ to 0.20~mJy~beam$^{-1}$.}\label{fig1}
\end{figure}

\placefigure{fig1}

\begin{figure}
\plotone{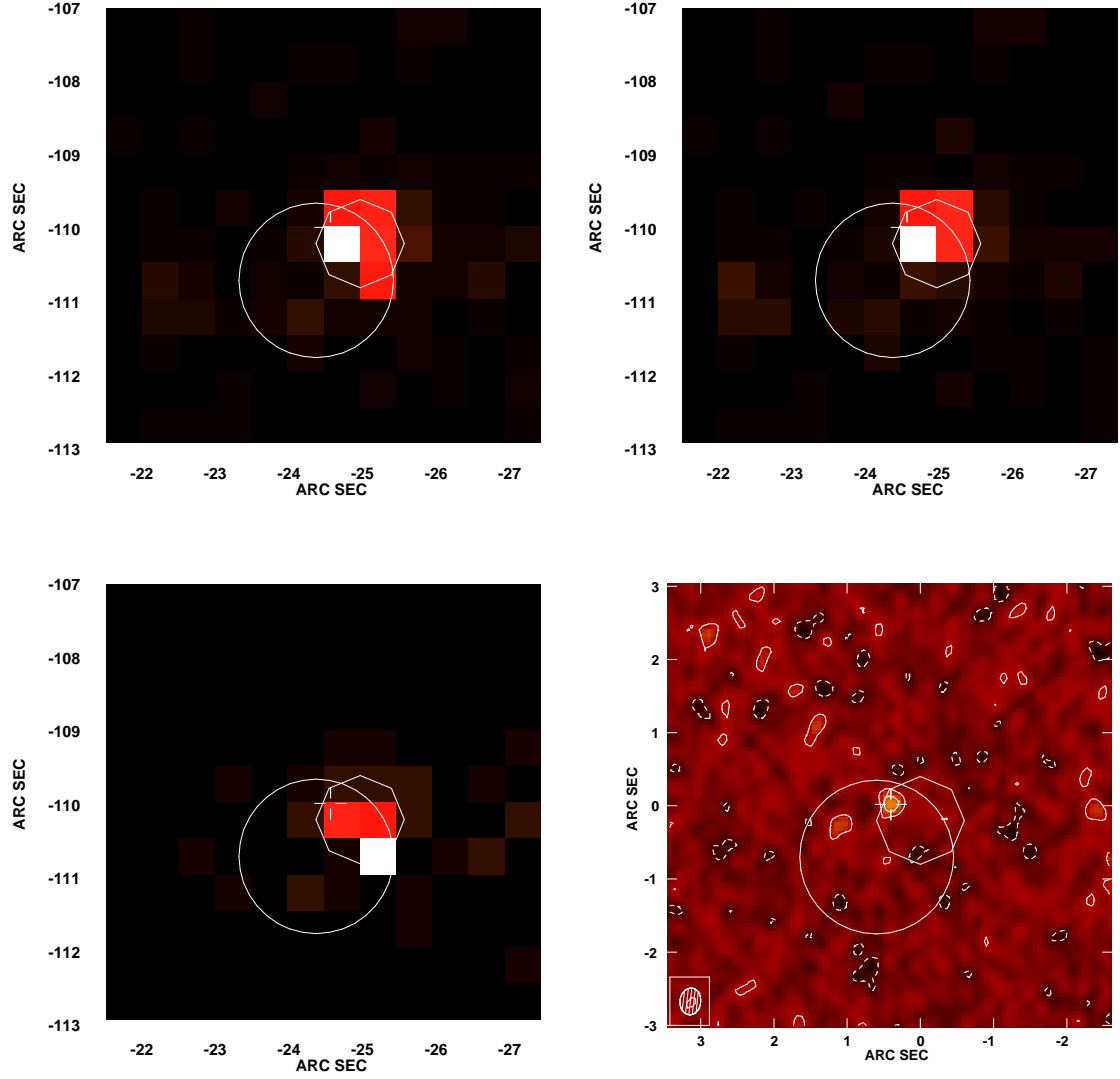}
\caption{Images of the central 6\arcsec\, (340 pc) of NGC\,4697.  In
the {\em Chandra\/} images the peaks are 18 counts {\em (top left)},
14 counts {\em (top right)\/} and 6 counts {\em (bottom left)}.  In
the VLA image the noise is 0.017~mJy~beam$^{-1}$ (1 $\sigma$) and the
beam dimensions at FWHM are 0.37\arcsec\, $\times$ 0.28\arcsec\, with
elongation PA = -6\arcdeg\, {\em (bottom right)}.  Otherwise as for
Figure~1.}\label{fig2}
\end{figure}

\placefigure{fig2}

Spectra were extracted from circular regions centered at the X-ray
positions of the candidate LLAGNs.  Extraction radii of 2.5 pixels
(1.2\arcsec) and 3 pixels (1.5\arcsec) were used for NGC\,4621 and
NGC\,4697, respectively, to maximize the signal-to-noise ratio and to
minimize the contribution from adjacent emission.  Spectra of the
background were taken from an off-nuclear region in the same field of
view and subtracted from the source spectra.  Each spectral bin
contains 12 and 10 counts for NGC\,4621 and NGC\,4697, respectively.
Thus a Gaussian approximation of a Poisson distribution is not
appropriate and a chi-squared method cannot be used for parameter
estimation.  Instead, the $C$ statistic, a likelihood defined by using
a Poisson distribution, was employed \citep{cas79}.  Spectral fits
were performed by using XSPEC version 11.3.  The spectra were fitted
with a power-law model modified by the total absorption along the line
of sight, thus including contributions both from the Galaxy and
intrinsic to the elliptical galaxies.  The best-fit parameters and
errors appear in Table~1.  Quoted errors are at the 90\% confidence
level for one parameter of interest.  Table~1 also gives the $C$
statistic per number of spectral bins, an indicator of the quality of
the fit.  The fitted absorption columns are consistent with the
Galactic values of 2.2 and 2.1 $\times 10^{20}$ cm$^{-2}$ for
NGC\,4621 and NGC\,4697, respectively \citep{dic90}.  The spectra and
best-fit models are shown in Figure~3.

\begin{figure}
\plotone{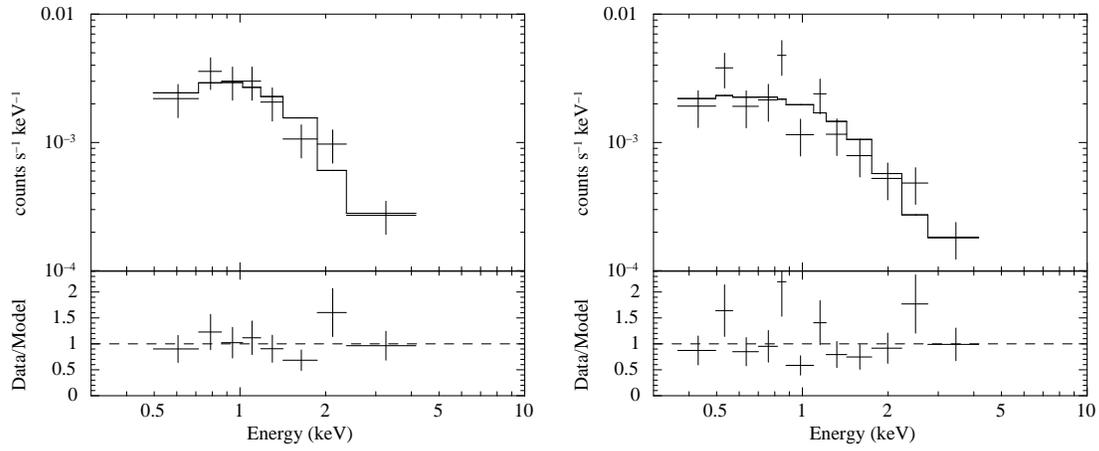}
\caption{{\em Chandra\/} spectra of the LLAGNs in NGC\,4621 {\em
(left)\/} and NGC\,4697 {\em (right)}.  Each top panel shows crosses
for the data and a histogram for the model.  Each bottom panel shows
the ratio of the data to the model, a preferred diagnostic when the
$C$-statistic is employed.}\label{fig3}
\end{figure}

\placefigure{fig3}

In addition to the candidate LLAGNs, Figures~1 and 2 show several
faint X-ray sources as well as hints of extended soft X-ray emission.
To provide further information about the extended emission and the
population of discrete sources in each galaxy, adaptively-smoothed
images were made over the central 3\arcmin\, in the soft (0.5-2~keV)
energy band.  These images appear in Figure~4.

\begin{figure}
\plottwo{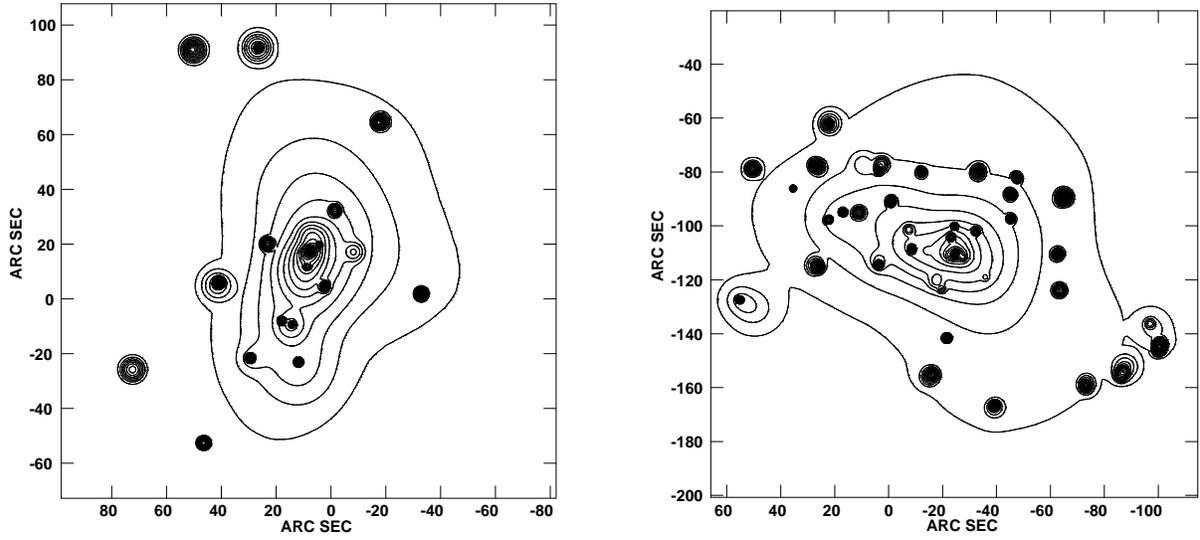}{f4b.eps}
\caption{Adaptively-smoothed {\em Chandra\/} images at 0.5-2 keV with
contours spaced by factors of the square root of 2.
{\em Left:\/} Central 3\arcmin\, (16 kpc) of NGC\,4621.  Lowest
contour is 0.007 and peak is 6.8, in units of smoothed counts.
{\em Right:\/} Central 3\arcmin\, (10 kpc) of NGC\,4697.  Lowest
contour is 0.02 and peak is 39, in units of smoothed
counts.}\label{fig4}
\end{figure}

\placefigure{fig4}

\section{New Data}\label{data-new}

Under proposal code AW671, the VLA was used in the A configuration to
observe NGC\,4621 at 8.5~GHz using phase calibrator J1254+1141 and
NGC\,4697 at 8.5~GHz using phase calibrator J1246-0730.  Phase
calibrator positions were taken from the Goddard VLBI global solution
2004 f,
\footnote{http://gemini.gsfc.nasa.gov/solutions/2004f\_astro/} and had
one-dimensional errors at 1 $\sigma$ better than 1 mas.  The switching
time between a galaxy and its phase calibrator was 460~s, while
switching angles were about 3\arcdeg\, or less.  The {\em a priori\/}
pointing positions for NGC\,4621 and NGC\,4697 were taken from
\citet{wro00b} and \citet{sar01}, respectively.  Data were acquired in
dual circular polarizations with a bandwidth of 100~MHz.  Observations
were made assuming a coordinate equinox of 2000.  The UT observation
date and net exposure times appear in Table~2.  Observations of
3C\,286 were used to set the amplitude scale to an accuracy of about
3\%.  Twenty-two of 27 antennas provided data of acceptable quality,
with most of the data loss attributable to EVLA retrofitting
activities.  The data were calibrated using the 2006 December 31
release of the NRAO AIPS software.  No self-calibrations were
performed.  No polarization calibration was performed, as only upper
limits on the galaxies' linear polarization percentages were sought.

The AIPS task {\tt imagr} was used to form and deconvolve images of
the Stokes $I\/$ emission at 8.5~GHz from each galaxy.  The images,
made with natural weighting to obtain the lowest 1 $\sigma$ noise
levels, appear in Figures~1 and 2.  Each galaxy image was searched
within the NED/2MASS error circle for emission above 4 $\sigma$,
leading to the detection of an unresolved source, with diameter less
that 0.3\arcsec.  Quadratic fits in the image plane yielded the peak
flux densities appearing in Table~1, along with their errors that are
quadratic sums of the 3\% scale error and $\sigma$.  Those fits also
yielded the radio positions listed in Table~2 along with their 90\%
position-errors obtained from the quadratic sum of a term due to the
signal-to-noise ratio of the detection and a term due to the
phase-referencing strategies, estimated to be $\sigma = 0.1\arcsec$
(the errors in the phase calibrator positions were negligible).  The
radio positions are marked in Figures~1 and 2.  {\tt imagr} was also
used to form naturally-weighted images of Stokes $Q\/$ and $U\/$ from
each galaxy.  Those images showed the same values for $\sigma$ as
their Stokes $I\/$ counterparts but led to no detections.

\section{Implications from the Astrometry}\label{astrometry}

Each of these nearby elliptical galaxies have been detected in the new
high-resolution VLA images shown in Figures~1 and 2.  Each
high-resolution radio detection occurs within the error circle for the
galaxy centroid position.  Thus each galaxy harbors a bona fide LLAGN
that has been localized at radio frequencies with subarcsecond
accuracy.  The position of the high-resolution radio detection of
NGC\,4621 is consistent with the previous radio detection at low
resolution, which was taken to mark a candidate LLAGN.  NGC\,4697 has
no prior radio detection and thus no prior radio astrometry.

Candidate LLAGNs have been identified in the high-resolution {\em
Chandra\/} images derived from archival data and shown in Figures~1
and 2.  The position of the candidate LLAGN in NGC\,4697 is consistent
with that reported for Source 1 by \citet{sar01}.  NGC\,4621 has no
prior X-ray astrometry.  Formally, the X-ray positions of these two
candidate LLAGNs are consistent with the centroid positions of the
host galaxies from NED/2MASS.

For NGC\,4621 and NGC\,4697, the separations between the
high-resolution radio and X-ray positions are about 0.6\arcsec\, and
0.5\arcsec, respectively.  The quadratic sum of the 1 $\sigma$ error
in the radio astrometry (0.1\arcsec) and the X-ray astrometry
(0.3\arcsec) is about 0.32\arcsec, leading to normalized separations
of $0.6\arcsec / 0.32\arcsec \sim 1.9$ for NGC\,4621 and $0.5\arcsec /
0.32\arcsec \sim 1.6$ for NGC\,4697.  These normalized separations are
less than 3, implying valid radio/X-ray matches for both galaxies
\citep{con78}.

Thus the X-ray positions of the candidate LLAGNs are consistent with
the positions of their high-resolution radio detections, meaning that
NGC\,4621 and NGC\,4697 each host a bona fide LLAGN that has been
identified at both X-ray and radio frequencies.  The X-ray sources are
therefore referred to as LLAGNs in Table~2 and Figure~3.

\section{Implications from the Photometry}\label{photometry}

\subsection{X-Ray Properties}

From Table~1, the X-ray luminosities of the LLAGNs in NGC\,4621 and
NGC\,4697 conform to the luminosity limit of $L(2-10~keV) <
10^{42}$~ergs~s$^{-1}$ for the X-ray definition of a LLAGN
\citep{ter03}.  The tabulated photon indices conform to the range of
values, 1.6 to 2.0, shown by other LLAGNs \citep{ter03}.  The X-ray
photometry given in Table~1 for the LLAGN in NGC\,4697 is consistent
with that reported for Source 1 by \citet{sar01} and \citet{sor06a}.
Also, the latter study analyses data from 2000 and from 2003-2004, and
finds no evidence for significant spectral or luminosity changes over
that long term.

NGC\,4621 and NGC\,4697 each host a supermassive black hole with
$M_\bullet > 10^8 M_\odot$, for which the characteristic Eddington
luminosity is $L(Edd) > 10^{46}$~ergs~s$^{-1}$ (Table~1).  In stark
contrast, the observed X-ray luminosities are lower by about 9 orders
of magnitude, and using them as a proxy for a bolometric luminosity,
the black holes energizing the LLAGNs in NGC\,4621 and NGC\,4697 are
found to have Eddington ratios $L(2-10~keV) / L(Edd) \sim 10^{-9}$
(Table~1).  Such ratios are very sub-Eddington.

\subsection{Radio and Circumnuclear Properties}

For the LLAGN in NGC\,4621, the flux density at 8.5~GHz measured at a
resolution of about 10\arcsec\, (880~pc) in 1999 (\S~\ref{data-old})
is higher than that measured at a resolution of 0.3\arcsec\, (26~pc)
in 2006 (Table~1).  This difference could arise from time variability
and/or from source resolution effects in NGC\,4621.  Variability on
typical timescales of a few days has been established at 8.5~GHz for
other LLAGNs \citep{and05}.  A difference due to source resolution
effects in NGC\,4621 could be linked to jet-like emission driven by
the LLAGN, or to the stellar substructures on scales of about
2\arcsec\, (180~pc) that are traced either kinematically \citep{wer02}
or photometrically \citep{kra04}.  But a stellar origin seems unlikely
for two reasons.  First, the Palomar spectrum shows only an old
stellar population \citep{ho03a}.  Second, if the flux-density
difference arises from star formation, then converting it to 1.4~GHz
with a spectral index of -0.7 implies a star-formation rate of about
0.0045 $M_\odot$~yr$^{-1}$ \citep{yun01}, for which the H$\alpha$
luminosity \citep{ken98} would be about 6 times higher than its
observed upper limit \citep{ho03a}.  The left panel of Figure~4 shows
a population of discrete X-ray sources, presumably low-mass X-ray
binaries.  There also appears to be soft X-ray emission on a scale of
a few kiloparsecs, but this faint emission may partly arise from
blends of discrete sources and, based on Figure~4, it will be
difficult to characterize the diffuse gas potentially available for
accretion onto the black hole.

For the LLAGN in NGC\,4697, the flux-density upper limit at 8.5~GHz
measured at a resolution of about 3\arcsec\, (170~pc) in 2000
(\S~\ref{data-old}) is consistent with the detection at a resolution
of 0.3\arcsec\, (17~pc) in 2006 (Table~1).  NGC\,4697 has a probable
stellar cluster centered on a dusty disk with a diameter of 7\arcsec\,
(400~pc) \citep{sor06a}.  The right panel of Figure~4 shows evidence
for soft X-ray emission on a scale of a few kiloparsecs, consistent
with prior reports by \citet{sar01} and \citet{sor06a}.  This diffuse
emission is characterized by a ``temperature'' of 0.33~keV and a
central gas density of 0.02~cm$^{-3}$; this gas, as well as the gas
released by the stellar populations, is potentially available for
accretion within the black hole's sphere of influence of diameter
0.76\arcsec\, (44~pc) \citep{sor06a,sor06b}.  The right panel of
Figure~4 also shows a population of discrete X-ray sources, identified
as low-mass X-ray binaries in prior studies \citep{sar01,sor06a}.

\subsection{Radio Loudness}

The study of Palomar LLAGNs using {\em Chandra\/}, the VLA, and the
VLBA found that the majority are radio loud, defined as $log~R_X =
log~\nu L_{\nu}(5~GHz) / L(2-10~keV) = -4.5$ or higher \citep{ter03}.
That study involved radio sources with flat or inverted spectra, so
the cited definition applies equally well at 8.5~GHz as at 5~GHz.  As
shown in Table~1, the emission from the quiescent black holes in
NGC\,4621 and NGC\,4697 is radio-loud, with $log~R_X = log~\nu
L_{\nu}(8.5~GHz) / L(2-10~keV) \sim -2$.  Since radio continuum
emission is outflow-dominated \citep{nag05}, this suggests the
presence of a radio outflow on scales less that 26~pc in NGC\,4621 and
less than 17~pc in NGC\,4697 (\S~\ref{data-new}).  Baring strong
projection effects, the length scale of the 8.5-GHz emission from
NGC\,4697 would place it inside the black hole's sphere of influence
\citep{sor06a,sor06b}.

\subsection{The Radio--X-ray--Mass Relation}

A relationship among radio luminosity, X-ray luminosity and black-hole
mass was discovered by \citet{mer03} and \citet{fal04} for
accretion-powered systems.  For LLAGNs in particular this so-called
fundamental plane (FP) relation supports the idea that they are
massive analogs of black hole X-ray binaries in the low/hard state,
with Eddington ratios of about $10^{-5}$ to $10^{-3}$.  Within this
framework, general models for the central engine of a LLAGN
\citep[e.g.,][]{yua02b} contain (1) a cool, optically-thick,
geometically-thin accretion disk with a truncated inner radius; (2) a
hot, radiatively-inefficient accretion flow (RIAF) interior to the
truncation radius; and (3) an outflow/jet.  Application of such models
to black hole X-ray binaries in the low/hard state can lead to strong
tests involving both spectral and timing properties.  For example, the
coupled accretion-jet model of \citet{yua05a} was successfully tested
in this way.  The outflow emission in that model was treated using the
internal-shock picture widely applied to gamma-ray bursts.  For the
low/hard binary tested, the outflow dominated the radio and infrared
emission, the thin disk dominated the UV emission, the hot RIAF
dominated the X-ray emission, and all three components contributed to
the optical emission.  In addition, the \citet{yua05a} model was shown
to be consistent with the FP relation \citep{yua05b}.  Cast in the
notation of Table~1, the FP relation of \citet{mer03} is $log \nu
L_\nu(5~GHz) = 0.60 log L(2-10 keV) + 0.78 log (M_\bullet/M_\odot) +
7.33$, with a scatter of $\sigma = 0.88$.  Given this large scatter,
the relation can be use to estimate the 8.5-GHz luminosity $\nu
L_\nu(8.5~GHz)$ expected for NGC\,4621 and NGC\,4697 from their black
hole masses and X-ray luminosities (Table~1).  These FP-based
predictions are $\nu L_\nu(8.5~GHz) \sim 4.0 \times 10^{36}$ ergs
s$^{-1}$ for NGC\,4621 and $\nu L_\nu(8.5~GHz) \sim 1.4 \times
10^{36}$ ergs s$^{-1}$ for NGC\,4697.  Although these predictions have
uncertainties of almost an order of magnitude, both are about that
amount below the galaxys' observed $\nu L_\nu(8.5~GHz)$ listed in
Table~1.

The black holes in NGC\,4621 and NGC\,4697 are very sub-Eddington,
featuring ratios of about $10^{-9}$ (Table~1).  For other similar
systems, departures from the FP relation have been suggested
\citep[e.g.,][]{fen03,mar05}.  Within the context of the coupled
accretion-jet model \citep{yua05a} described above, the very
sub-Eddington ratios for NGC\,4621 and NGC\,4697 place them in the
model's so-called quiescent regime \citep{yua05b}.  Moreover, the
black holes in NGC\,4621 and NGC\,4697 are able to generate radio
emission which, as argued in the previous subsection, is plausibly
jet-like.  Thus the ``if the jet persists'' caveat for the quiescent
regime seems to be fulfulled, and motivates application of the
radio--X-ray--mass relation for the quiescent regime \citep{yua05b}.
Cast in the notation of Table~1, that quiescent relation is $log \nu
L_\nu(8.5~GHz) = 1.23 log L(2-10 keV) + 0.25 log (M_\bullet/M_\odot) -
13.45$, and the black hole masses and X-ray luminosities in Table~1
yield the predicted radio luminosities $\nu L_{\nu}(8.5~GHz)$
appearing in Table~1.  Those predicted radio luminosities agree, to
within a factor of a few, with the observed radio luminosities, also
listed in Table~1.  Significantly, the radio--X-ray--mass relation for
the quiescent regime invokes X-ray emission from the outflow rather
than from an accretion flow.  Also, \citet{yua05b} note that their
index for the quiescent relation, 1.23, is in general agreement with
earlier pure outflow models \citep[e.g.,][]{hei04}.

In the coupled accretion-jet model \citep{yua05a}, the outflow/jet
generates synchrotron emission which is optically-thick near 8.5~GHz
and optically-thin at hard X-rays.  In the model's quiescent regime
\citep{yua05b}, the emission from the outflow dominates over that from
any thin disk and any hot RIAF.  Specifically, the spectral energy
distribution of the outflow curves smoothly across the electromagnetic
spectrum, and photon indices near 2, like those observed (Table~1),
seem achievable in the hard X-rays \citep{yua05b}.  For comparison,
Comptonization in a RIAF acts as a natural thermostat, limiting the
electron temperature to about 100~keV.  At the very low accretion
rates associated with the quiescent regime, any feeble RIAF would
generate 100-keV bremsstrahlung emission which has a photon index of
1.3 in the {\em Chandra\/} band.  Such an index is not consistent with
the measured value for NGC\,4621 but is just consistent with the
measured value for NGC\,4697 (Table~1).  In the latter case, some
bremsstrahlung contribution to the 2-10~keV luminosity could account
for the model's slight overprediction of the 8.5-GHz luminosity.

\section{Summary and Future Directions}\label{summary}

Analysis of {\em Chandra\/} data and VLA data shows that the nearby
elliptical galaxies NGC\,4621 and NGC\,4697 each contain a
low-luminosity active galactic nucleus (LLAGN), identified as a faint,
hard X-ray source that is astrometrically coincident with a faint
8.5-GHz source.  These frequency regions benefit from high contrast
against the stellar emission, little or no obscuration, and
subarcsecond angular resolution and astrometry.

The massive black holes energizing these LLAGNs have Eddington ratios
$L(2-10~keV) / L(Edd) \sim 10^{-9}$, placing them in the quiescent
regime.  These quiescent black holes are radio-loud emitters,
suggesting the presence of a radio outflow.  Application of the
radio--X-ray--mass relation from Yuan \& Cui for quiescent black holes
predicts the observed radio luminosiites $\nu L_{\nu}(8.5~GHz)$ to
within a factor of a few.  Importantly, that relation invokes X-ray
emission from the outflow rather than from an accretion flow.  In the
radio and X-ray regions, the faint, but detectable, emission from
these two massive black holes is therefore consistent with being
outflow-dominated.

Clearly, the spectral energy distributions of the LLAGNs in NGC\,4621
and NGC\,4697 need to be measured across the electromagnetic spectrum.
Deep VLA observations at frequencies of 1-50 GHz would test the Yuan
\& Cui prediction of optically-thick emission from the LLAGNs, as well
as enable searches for faint, jet-like structures adjacent to the
LLAGNs.  Also, short-term variability on timescales of a few days has
been found at 8.5~GHz for other LLAGNs \citep{and05}.  For each of the
LLAGNs in NGC\,4621 and NGC\,4697, evidence for short-term, correlated
variability between the radio source and the X-ray source would
strengthen the case for their common, outflow origin.  Finally, more
cases like NGC\,4621 and NGC\,4697 are needed to better define the
population of quiescent LLAGNs, and the improved sensitivity of the
Expanded VLA will help enormously on this front \citep{ulv06}.

\acknowledgments We acknowledge the valuable feedback from an
anonymous referee.

{\it Facilities:} \facility{{\em Chandra}}, \facility{VLA}.

\clearpage

\clearpage

\clearpage

\clearpage

\clearpage

\begin{deluxetable}{lcccc}
\tabletypesize{\scriptsize}
\tablecolumns{5}
\tablewidth{0pc}
\tablecaption{Parameters of the LLAGNs}\label{tab1}
\tablehead{
\colhead{Parameter} & \colhead{NGC\,4621} & \colhead{Ref.} &
\colhead{NGC\,4697} & \colhead{Ref.}}
\startdata
$D$ (Mpc)                                  & 18.2   & 1 & 11.7   & 2 \\
$s$ (pc arcsec$^{-1}$)                     & 88     & 1 & 57     & 2 \\
$M_\bullet$ ($10^8 M_\odot$)               & 2.7    & 3 & 1.7    & 2 \\   
$L(Edd)$ ($10^{46}$ ergs s$^{-1}$)         & 3.5    & 3 & 2.2    & 2 \\
$N_{\rm H}$ ($10^{20}$ cm$^{-2}$)          & $<$18. & 4 & $<$8.4 & 4 \\
$\Gamma$         & 1.8$^{+0.8}_{-0.3}$ & 4 & 1.6$^{+0.5}_{-0.3}$ & 4 \\
$C$ statistic per number of spectral bins  & 5/8    & 4 & 17/12  & 4 \\
$F(2-10~keV)$ ($10^{-14}$ ergs s$^{-1}$ cm$^{-2}$)  & 
                           2.1$^{+1.2}_{-1.0}$ & 4 & 1.7$\pm$0.7 & 4 \\
$L(2-10~keV)$ ($10^{37}$ ergs s$^{-1}$)    & 6.6    & 4 & 2.2    & 4 \\
$L(2-10~keV) / L(Edd)$ ($10^{-9}$)         & 1.9    & 4 & 1.0    & 4 \\
$S(8.5~GHz)$ (mJy)       & 0.098$\pm$0.018 & 4 & 0.092$\pm$0.017 & 4 \\
Observed $\nu L_{\nu}(8.5~GHz)$ ($10^{35}$ ergs s$^{-1}$)  
 & 3.3 & 4 & 1.3 & 4 \\
$log~R_X=log~\nu L_{\nu}(8.5~GHz)/L(2-10~keV)$ & -2.3 & 4 & -2.2 & 4 \\
Predicted $\nu L_{\nu}(8.5~GHz)$ ($10^{35}$ ergs s$^{-1}$)
 & 1.5 & 5 & 3.5 & 5 \\
\enddata
\tablecomments{
Row~(1): Surface-brightness-fluctuation distance.
Row~(2): Scale.
Row~(3): Mass of black hole.
Row~(4): Eddington luminosity of black hole.
Row~(5): Galactic plus intrinsic column density.
Row~(6): Photon index.
Row~(7): $C$ statistic (Cash 1979).
Row~(8): 2-10 keV flux.
Row~(9): 2-10 keV luminosity.
Row~(10): Eddington ratio.
Row~(11): 8.5~GHz flux density.
Row~(12): Observed radio luminosity.
Row~(13): Radio loudness.
Row~(14): Predicted radio luminosity.}
\tablerefs{(1) Ravindranath et al. 2002; (2) Pellegrini 2005; (3)
Tremaine et al. 2002; (4) this work; (5) Yuan \& Cui 2005}
\end{deluxetable}
\clearpage

\begin{deluxetable}{llcllcccc}
\tabletypesize{\scriptsize}
\tablecolumns{9}
\tablewidth{0pc}
\tablecaption{Astrometry of the Galaxy Components}\label{tab2}
\tablehead{
\colhead{}          & \colhead{}          & \colhead{}          &
\colhead{R.A.}      & \colhead{Decl.}     & \colhead{Error}     &
\colhead{}          & \colhead{Exposure}  & \colhead{}          \\
\colhead{Galaxy}    & \colhead{Component} & \colhead{Region}    &
\colhead{(J2000)}   & \colhead{(J2000)}   & \colhead{(\arcsec)} &
\colhead{Date}      & \colhead{(s)}       & \colhead{Ref.}      \\
\colhead{(1)}       & \colhead{(2)}       & \colhead{(3)}       &
\colhead{(4)}       & \colhead{(5)}       & \colhead{(6)}       &
\colhead{(7)}       & \colhead{(8)}       & \colhead{(9)}       }
\startdata
NGC\,4621 ... 
 & Galaxy centroid 
   &K& 12 42 02.32  & 11 38 48.9  & 2.1   & \nodata      & \nodata & 1\\
 & LLAGN           
   &X& 12 42 02.25  & 11 38 49.5  & 1.2   & 2001 Aug 1     & 24837 & 2\\
 & LLAGN           
   &R& 12 42 02.256 & 11 38 48.93 & 0.43  & 2006 Apr 21    &  6630 & 2\\
NGC\,4697 ... 
 & Galaxy centroid 
   &K& 12 48 35.91  & -05 48 03.1  & 2.1  & \nodata      & \nodata & 1\\
 & LLAGN           
   &X& 12 48 35.87  & -05 48 02.6  & 1.2  & 2000 Jan 15-16 & 39260 & 2\\
 & LLAGN           
   &R& 12 48 35.897 & -05 48 02.38 & 0.43 & 2006 Apr 21    &  6060 & 2\\
\enddata
\tablecomments{
Col.~(1): Galaxy name. 
Col.~(2): Nuclear component.
Col.~(3): Frequency region coded as K for near-infrared, R for radio
at 8.5~GHz and X for X-ray.
Cols.~(4) and (5): Component position.  Units of right ascension are 
hours, minutes, and seconds, and units of declination are degrees, 
arcminutes, and arcseconds.
Col.~(6): Diameter of error circle at 90\% confidence level.
Col.~(7): UT observation date.
Col.~(8): Exposure time.
Col.~(9): Reference.}
\tablerefs{(1) NED/2MASS; (2) this work.}
\end{deluxetable}
\clearpage

\begin{thebibliography}{}
\bibitem[Anderson \& Ulvestad(2005)]{and05} Anderson, J. M., \&
 Ulvestad, J. S. 2005, \apj, 627, 674
\bibitem[Bietenholz et al.(2000)]{bie00} Bietenholz, M. F., Bartel,
 N., \& Rupen, M. P. 2000, \apj, 532, 895
\bibitem[Birkinshaw \& Davies(1985)]{bir85} Birkinshaw, M., \& Davies,
 R. L. 1985, \apj, 291, 32
\bibitem[Cash(1979)]{cas79} Cash, W. 1979, \apj, 228, 939
\bibitem[Condon \& Dressel(1978)]{con78} Condon, J. J., \& Dressel,
 L. L. 1978, \apj, 221, 456
\bibitem[Condon et al.(1998)]{con98} Condon, J. J., Cotton, W. D.,
 Greisen, E. W., Yin, Q. F., Perley, R. A., Taylor, G. B., \&
 Broderick, J. J. 1998, \aj, 115, 1693
\bibitem[Dickey \& Lockman(1990)]{dic90} Dickey, J. M., \& Lockman,
 F. J. 1990, \araa, 28, 215
\bibitem[Falcke \& Biermann(1999)]{fal99} Falcke, H., \& Biermann,
 P. L. 1999, \aap, 342, 49
\bibitem[Falcke \& Markoff(2000)]{fal00} Falcke, H., \& Markoff, S.
 2000, \aap, 362, 113
\bibitem[Falcke et al.(2004)]{fal04} Falcke, H., Koerding, E., \&
 Markoff, S. 2004, \aap, 414, 895
\bibitem[Fender et al.(2003)]{fen03} Fender, R. P., Gallo, E., \&
 Jonker, P. G. 2003, \mnras, 343, L99
\bibitem[Heinz(2004)]{hei04} Heinz, S. 2004, \mnras, 355, 835
\bibitem[Ho(2004)]{ho04} Ho, L. C.. 2004, Coevolution of Black Holes
 and Galaxies, ed. L. C. Ho (Cambridge: CUP), 292
\bibitem[Ho et al.(1997)]{ho97} Ho, L. C., Filippenko, A. V., \&
 Sargent, W. L. W. 1997, \apjs, 112, 315
\bibitem[Ho et al.(2001)]{ho01} Ho, L. C., et al. 2001, \apj, 549, L51
\bibitem[Ho et al.(2003a)]{ho03a} Ho, L. C., Filippenko, A. V., \& 
 Sargent, W. L. W. 2003a, \apj, 583, 159
\bibitem[Ho et al.(2003b)]{ho03b} Ho, L. C., Terashima, Y., \&
 Ulvestad, J. S.  2003b, \apj, 589, 783
\bibitem[Kennicutt(1998)]{ken98} Kennicutt, R. C., Jr. 1998, \araa,
 36, 189
\bibitem[Kormendy(2004)]{kor04} Kormendy, J. 2004, Coevolution of
 Black Holes and Galaxies, ed. L. C. Ho (Cambridge: CUP), 1
\bibitem[Krajnovic \& Jaffe(2002)]{kra02} Krajnovic, D., \& Jaffe,
 W. 2002, \aap, 390, 423
\bibitem[Krajnovic \& Jaffe(2004)]{kra04} Krajnovic, D., \& Jaffe,
 W. 2004, \aap, 428, 877
\bibitem[Livio et al.(1999)]{liv99} Livio, M., Ogilvie, G. I., \&
 Pringle, J. E. 1999, \apj, 512, 100
\bibitem[Markoff(2005)]{mar05} Markoff, S. 2005, \apj, 618, L103
\bibitem[Meier(2001)]{mei01} Meier, D. L. 2001, \apj, 548, L9
\bibitem[Merloni et al.(2003)]{mer03} Merloni, A., Heinz, S., \&
 Di Matteo, T. 2003, \mnras, 345, 1057
\bibitem[Nagar et al.(2005)]{nag05} Nagar, N. M., Falcke, H., \& Wilson,
 A. S. 2005, \aap, 435, 521
\bibitem[Narayan et al.(1995)]{nar95} Narayan, R., Yi, I., Mahadevan,
 R. 1995, \nat, 374, 623
\bibitem[Pellegrini(2005)]{pel05} Pellegrini, S. 2005, \apj, 624, 
 155
\bibitem[Pinkney et al.(2005)]{pin05} Pinkney, J., et al. 2005, \apj,
 596, 903
\bibitem[Ravindranath et al.(2002)]{rav02} Ravindranath, S., Ho,
 L. C., \& Filippenko, A. V.\ 2002, \apj, 566, 801
\bibitem[Sarazin et al.(2001)]{sar01} Sarazin, C. L., Irwin, J. A., \&
 Bregman, J. M. 2001, \apj, 556, 533
\bibitem[Soria et al.(2006a)]{sor06a} Soria, R., et al. 2006a, \apj,
 640, 126
\bibitem[Soria et al.(2006b)]{sor06b} Soria, R., et al. 2006b, \apj,
 640, 143
\bibitem[Terashima \& Wilson(2003)]{ter03} Terashima, Y., \& Wilson,
 A. S. 2003, \apj, 583, 145
\bibitem[Terashima et al.(2005)]{ter05} Terashima, Y., Ho, L. C., \&
 Ulvestad, J. S. 2005, The Interplay among Black Holes, Stars and ISM
 in Galactic Nuclei, IAUS 222, eds. T. Storchi-Bergmann et al.
 (Cambridge: CUP), 61
\bibitem[Thompson et al.(1980)]{tho80} Thompson, A. R., Clark, B. G.,
 Wade, C. M., Napier, P. J., 1980, \apjs, 44, 151
\bibitem[Tremaine et al.(2002)]{tre02} Tremaine, S., et al. 2002,
 \apj, 574, 740
\bibitem[Ulvestad et al.(2006)]{ulv06} Ulvestad, J. S., Perley, R. A.,
 McKinnon, M. M., Owen, F. N., Dewdney, P. E., \& Rodriguez, L. F.
 2006, \baas, 38, 135
\bibitem[Wernli et al.(2002)]{wer02} Wernli, F., Emsellem, E., \& 
 Copin, Y. \aap, 396, 73
\bibitem[White et al.(1997)]{whi97} White, R. L., Becker, R. H.,
 Helfand, D. J., \& Gregg, M. D.  1997, \apj, 475, 479
\bibitem[Wrobel \& Heeschen(1991)]{wro91} Wrobel, J. M., \& Heeschen,
 D. S.  1991, \aj, 101, 148
\bibitem[Wrobel \& Herrnstein(2000)]{wro00b} Wrobel, J. M., \&
 Herrnstein, J. R. 2000, \apj, 533, L111
\bibitem[Yuan et al.(2000a)]{yua02a} Yuan, F., Markoff, S., \& Falcke,
 H. 2000a, \aap, 383, 854
\bibitem[Yuan et al.(2000b)]{yua02b} Yuan, F., Markoff, S., Falcke,
 H., \& Biermann, P. L. 2000b, \aap, 391, 139
\bibitem[Yuan et al.(2005)]{yua05a} Yuan, F., Cui, W., \& Narayan, R.
 2005, \apj, 620, 905
\bibitem[Yuan \& Cui(2005)]{yua05b} Yuan, F. \& Cui, W. 2005, \apj,
 629, 408
\bibitem[Yun et al.(2001)]{yun01} Yun, M. S., Reddy, N. A., \& Condon,
 J. J. 2001, \apj, 554, 803
\end{thebibliography}
\end{document}